# Data Distillation for Neural Network Potentials toward Foundational Dataset

**Gang Seob Jung** [*]
Computational Sciences and Engineering Division
Oak Ridge National Laboratory
Oak Ridge, TN 37831
`jungg@ornl.gov`

**Sangkeun Lee**
Computational Sciences and Mathematics Division
Oak Ridge National Laboratory
Oak Ridge, TN 37831
`lees4@ornl.gov`

**Jong Youl Choi**
Computational Sciences and Mathematics Division
Oak Ridge National Laboratory
Oak Ridge, TN 37831
`choij@ornl.gov`

## Abstract

Machine learning (ML) techniques and atomistic modeling have rapidly transformed materials design and discovery. Specifically, generative models can swiftly propose promising materials for targeted applications. However, the predicted properties of materials through the generative models often do not match with calculated properties through *ab initio* calculations. This discrepancy can arise because the generated coordinates are not fully relaxed, whereas the many properties are derived from relaxed structures. Neural network-based potentials (NNPs) can expedite the process by providing relaxed structures from the initially generated ones. Nevertheless, acquiring data to train NNPs for this purpose can be extremely challenging as it needs to encompass previously unknown structures. This study utilized extended ensemble molecular dynamics (MD) to secure a broad range of liquid- and solid-phase configurations in one of the metallic systems, nickel. Then, we could significantly reduce them through active learning without losing much accuracy. We found that the NNP trained from the distilled data could predict different energy-minimized closed-pack crystal structures even though those structures were not explicitly part of the initial data. Furthermore, the data can be translated to other metallic systems (aluminum and niobium), without repeating the sampling and distillation processes. Our approach to data acquisition and distillation has demonstrated the potential to expedite NNP development and enhance materials design and discovery by integrating generative models.

---

[*]Corresponding Author

[1]Notice: This manuscript has been authored by UT-Battelle, LLC, under Contract No. DE-AC05-00OR22725 with the U.S. Department of Energy. The United States Government retains and the publisher, by accepting the article for publication, acknowledges that the United States Government retains a nonexclusive, paid-up, irrevocable, world-wide license to publish or reproduce the published form of this manuscript, or allow others to do so,

Preprint. Under review.

# 1 Introduction

Traditional approaches to materials design involve probing the physical and/or chemical properties of possible candidates with time-consuming and expensive trial-and-error studies. Therefore, deriving the principles for rational design from them is obstructed by the lack of sufficient and systematic data. Recent advances in physics-based computational atomistic modeling and simulations and machine learning (ML) techniques provide a novel avenue for exploring hypothetical materials to narrow the design space for materials with desired target properties [1, 2].

ML techniques have drastically advanced various sciences and engineering [3, 4, 5, 6, 7, 8]. Screening materials candidates through computational models with ML can accelerate new materials development and design, such as alloy [9], drugs [10], and polymer/protein engineering [11]. Although there are many challenges, from computational discovery to actual synthesis and deployment of the designed materials, the discovery is a critical starting point [12, 13].

To advance this computation-driven discovery as a starting point, two critical challenges must be addressed: the accuracy of ML models for predicting material properties and the effective generation of hypothetical materials. Although previous ML models work well for predicting the properties that are less sensitive to atomic coordinates, e.g., drug-likeness, it often becomes challenging to predict coordinate-sensitive properties, such as homo-lumo gap and vibration spectrum. This issue can be more severe with inorganic materials represented in the unit cell because their properties, such as band gap and elasticity, are more sensitive to the specific coordinates of atoms [14, 15, 16]. Therefore, obtaining the relaxed structure and the properties through *ab initio* calculations to confirm the prediction is often necessary. This redundant process can hinder the materials discovery. In many cases, the best properties predicted by the generated structures are less likely to be the same values after structural relaxation.

Recently developed ML-based forcefields (MLFFs) [17, 18, 19] have demonstrated a capability to predict potential energy surfaces (PESs) for atomic configurations with an accuracy comparable to *ab initio* electronic structure methods, but at a speed that is several orders of magnitude faster [20]. Neural network potentials (NNPs), a kind of MLFFs, utilize neural network to fit the interatomic interaction energies [21]. These have rapidly emerged due to their flexibility, accuracy, and efficiency. They are more suitable for large and complex systems than other MLFFs, as they can handle large data sets with many training data points.

Once NNPs can relax structure with similar accuracy as *ab initio* calculations, they can advance the materials discovery and screening more efficiently. However, NNPs, like other MLFFs, usually perform poorly outside their training domain and typically fail to predict unseen structures without appropriate data. Active learning (AL) [22] can help in improving the accuracy and exploring new structural data not included in the first stage when combining NNPs with enhanced sampling methods [23]. However, continuously acquiring more data and re-training still incurs significant computational costs. Therefore, it is desirable to have data in the beginning to handle the hypothetical structures as much as possible. Then, a smaller number of AL iterations are required for sufficient performance.

This study explores the feasibility of obtaining such "foundational" datasets for metallic systems using extended ensemble molecular dynamics (MD) techniques and data distillation. In MD simulations, conventional data sampling, e.g., isobaric-isothermal ensemble, has the local energy-minimum problem where sampling is more likely trapped in metastable or initial states. Therefore, it is challenging to sample amorphous structures or transient structures sufficiently. So many previous studies rely on changing temperatures from multiple known-initial structures to obtain various configurations for metallic/alloy systems [24, 25, 26]. In this study, we utilized the extended version of the multicanonical ensemble method [27, 28, 29], called the multiorder-multithermal ensemble (MOMT) [29] method. This method can sample possible transition states between solid and liquid phases with an accurate estimation of the density of states [30]. Also, we utilized the empirical embedded atom model (EAM) potentials [31, 32] instead of density functional theory calculations to evaluate the quality of sampled configuration.

Many similar configurations are sampled during either conventional or MOMT ensemble MD simulations. This can hinder the generalization of rarely sampled configurations. Therefore, based on the predicted atomic energy, we developed a data distillation process for the metallic system. Only 5% of the local configurational information can accurately describe the sampled structures (∼1.0 meV/atom). Also, the derived NNP can describe relaxed FCC, BCC, and HCP structures with the correct cohesive energy rank, although those are not explicitly included in the training. Furthermore, we successfully demonstrated the distilled configurations can be translated to another metallic system without repeating the sampling and distillation processes.

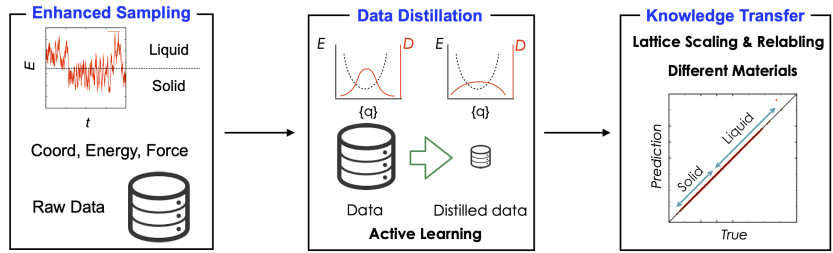

Figure 1: Schematic of workflow in the current study. First, we obtain the data of nickel (Ni) for two phases through the MOMT ensemble MD. The sampled data is biased to the equilibrium states. We reduce the amount of data without losing significant accuracy. The obtained knowledge on the configurations from Ni is translated to aluminum (Al) by scaling the lattice and relabeling.

## 2 Result and Discussion

A schematic of the workflow in the current study is shown in Figure 1. In the initial step, we sampled the configurations through the MOMT ensemble MD for liquid and solid phases. We utilized the previously developed scheme [30] to update the non-Boltzman weight factor for a broad sampling in both enthalpy and order parameter spaces. (See Supplementary Note 1). The method builds a bias for the visited states and allows overcoming the energy barrier to transit to other states. This approach is similar to well-known meta-dynamics or adaptive biasing force methods [33, 34]. One of the key differences of the multicanonical ensemble is that it utilizes the energy or enthalpy as a collective variable. Consequently, it can sample configurations with multiple reference temperatures. The difference in energy between the conventional MD and MOMT MD sampling is shown in Figure S1. Figure S1a shows the results of liquid and solid phases (108 nickel atoms) at the 2000K and 1 bar through the conventional isobaric-isothermal ensemble MD. Conventional sampling cannot overcome the energy barrier between the solid and liquid phases. However, the MOMT ensemble MD can sample both solid and liquid phases under the same conditions.

In this study, we utilized ANI-type NNP (See Supplementary Note 2). The results of the performance of NNPs through differently sampled configurations are summarized in Figure S2. The results confirm the previous findings that the training error (e.g., mean absolute error, MAE) values of training/validation data are insufficient to evaluate the NNPs because the values change because of configurational coverages [23, 35]. Although the data set from solid looks the best based on the training/validation MAE values. The MOMT sampled data demonstrates the best performance considering all sampled data. One notable fact is that even liquid phase-based NNP still decently predicts the solid phase while solid phase-based NNP poorly performs on the liquid data set.

Then, we evaluated the reliability of the trained NNPs by testing whether they can obtain relaxed FCC, BCC, and HCP structures through energy minimizations (See Supplementary Note 3). Table 1 shows the results. Although the FCC structure is the most stable state, only the NNP trained by the MOMT data could predict it correctly. However, even the NNP could not perfectly predict the other structures (BCC and HCP). A wider range of sampling could improve the reliability. Therefore, we sampled more to obtain 20,000 data points (Figure S3a). However, the method is inevitably sampling configurations of more probable states, not to visit next time by putting more bias and the sampled configurations can share similarity.

Therefore, we utilized active learning for data distillation to alleviate the data imbalance and generalize training NNPs. We performed 10 iterations for data distillation by the previously developed code [36]. We utilized the 5 models with randomly selected training (20%)/validation (80%) to estimate the atomic UQ from the standard deviation of the atomic energy predictions (See Supplementary Note 4). The purpose of the process is to reduce the amount of data without losing performance. An appropriate data reduction can help with fast training and performance, as shown in the previous study [35]. We chose 108 atom system as a reference because this is the minimum number to have a defective solid phase during the sampling, such as a stacking fault or HCP phase.

| Structure (# atoms) | Ni-FCC (32) $l_b$(Å); $E_{tot}$/atom (eV) | Ni-BCC (54) $l_b$(Å); $E_{tot}$/atom (eV) | Ni-HCP (48) $l_b$(Å); $E_{tot}$/atom (eV) |
|---|---|---|---|
| Reference (EAM) | 2.49/-4.876 | 2.41/-4.833 | 2.44/-4.847 |
| FCC Data | 2.56/-4.872 | 2.41/-4.905 | 2.52/-4.831 |
| (errors) | (+0.07/+0.004) | (0.00/-0.072) | (+0.08/+0.016) |
| LIQ Data | 2.43/-4.889 | 2.49/-4.793 | 2.45/-4.905 |
| (errors) | (-0.06/-0.013) | (+0.08/+0.040) | (+0.01/-0.058) |
| MOMT Data | 2.52/-4.895 | 2.46/-4.891 | 2.49/-4.889 |
| (errors) | (+0.03/-0.019) | (+0.05/-0.058) | (+0.05/-0.042) |

Table 1: Results of energy minimization from the different initial structures, FCC, BCC, and HCP nickel through NNPs with 2,000 data points (blue: lowest energy, red: highest energy, green: middle, lb: bond length). The self-energy values utilized for each model are listed in Table S4.

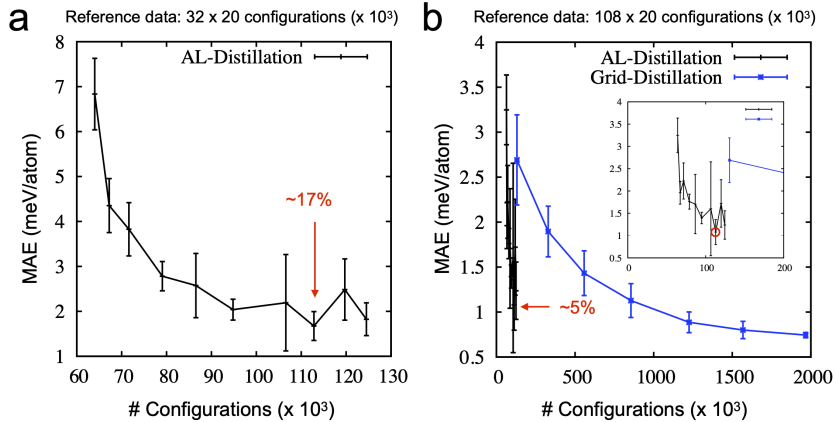

Figure 2: (a) The performance of data distillation through active learning for 32 nickel atoms system. The 17% data point can achieve MAE below 2.0 me/atom. The reference data for the MAE is MOMT sampled 20,000 frames (∼32×20,000 atomic configurations) (b) Comparison between grid-based distillation from 108 nickel atoms system and active learning-based distillation from 32 nickel atoms system. Only 5% of configurations can achieve MAE close to 1.0 meV/atom.

However, we found that the MOMT sampled configuration with a smaller unit cell of 32 atoms performs better as distilled data. Therefore, we used distilled data from the 32 atoms system. Figure 2a shows a good performance of data distillation on 32 nickel atoms system. 17% of configurations can achieve MAE lower than 2.0meV/atom for the total configurations. The effect is more drastic when we compare a physical more reasonable system (108 nickel atoms system) with manually distilled configurations in Figure 2b. The NNP trained from the distilled data shows a great performance (∼1.0meV/atom) with only 5% of information. As we hypothesized, widely sampled configurations can improve the energy rank of relaxed structures as shown in Table 2. The relaxed FCC, BCC, and HCP structures are well-matched with those referenced from EAM potentials (See Supplementary Note 5 for the performance between distilled onr and non-distilled one).

Finally, we investigated the generalization of the distilled information with the aluminum. Nickel and aluminum are FCC structures with different physical features regarding their energy, lattice parameters, and melting points. Aluminum has a larger lattice parameter (∼ 4.05Å) than nickel (∼3.52Å) and melt at a lower temperature (∼660°C) than nickel (1455°C). Although their lattice parameters are different, we can utilize the information by scaling the system size from nickel to aluminum (4.05/3.52∼1.15) and recalculating energy and forces based on aluminum potential. We also performed MOMT ensemble MD simulations with aluminum with 108 atoms to obtain 20,000 configurational data points (Figure S3b). The data was not utilized to train NNP but to evaluate the NNP trained by the translated data from the nickel system.

| Structure (#atoms) | Ni-FCC (32) $l_b$(Å); $E_{tot}$/atom (eV) | Ni-BCC (54) $l_b$(Å); $E_{tot}$/atom (eV) | Ni-HCP (48) $l_b$(Å); $E_{tot}$/atom (eV) |
|---|---|---|---|
| Reference (EAM) | 2.49/-4.876 | 2.41/-4.833 | 2.44/-4.847 |
| NNP (Distilled Data) | 2.50/-4.875 | 2.41/-4.812 | 2.46/-4.835 |
| (errors) | (+0.01/+0.001) | (0.00/+0.021) | (+0.02/0.012) |

Table 2: Results of energy minimization from the different initial structures, FCC, BCC, and HCP nickel through NNPs from distilled data. (blue: lowest energy, red: highest energy, green: middle, $l_b$: bond length)

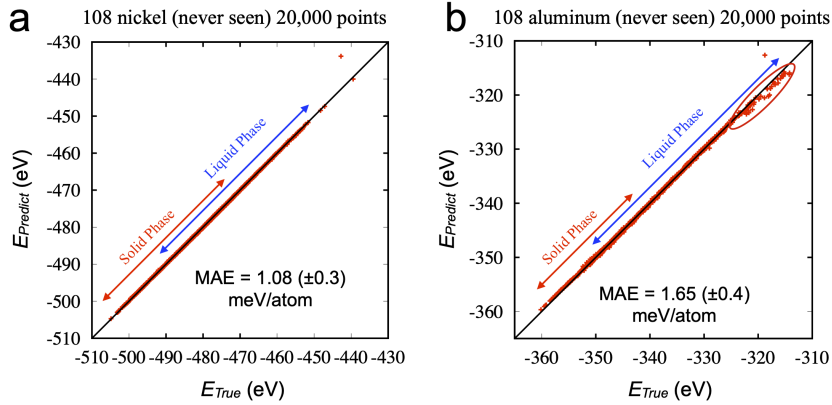

Figure 3: Scattering plot of the true and predicted energy values for 20,000 data points of 108 nickel atoms (1,000 points sampled for visualization). (b) Scattering plot of the true and predicted energy values for 20,000 data points of 108 aluminum atoms (1,000 points sampled for visualization). A red circle indicates the lower accuracy region, it accounts for rarely sampled energy region in the nickel system.

Figure 3 shows the scatter plot of both nickel and aluminum. The NNP for aluminum shows a high accuracy except for a high-energy region. The NNP can also describe the rank of three closed-pack crystals, as shown in Table 3. The selected data through the nickel system are our essential knowledge for the reliable NNPs to describe solid and liquid phases. Then, this knowledge could be well-translated to the other metallic systems, aluminum (See Supplementary Note 6 for BCC-Niobium case).

## 3 Conclusions

In this short report, we demonstrate the effectiveness of the MOMT ensemble sampling for generating training data for metallic systems. The sampled data from MOMT includes both solid and liquid phases and covers a wider range of energy than conventional sampling, such as the isothermal-isobaric ensemble. Data imbalance issues exist because the MD sampling inevitably collects more configurations near the equilibrium state. Utilizing the active learning process based on atomic UQ, we successfully distilled the data to 5% without losing much accuracy. Furthermore, we investigated the obtained configurations that can be utilized for a very different system, aluminum by simple scaling and relabeling. Our results promise that there can be foundational data for these systems' potential energy surface (PES). In the future, we will explore the data set with DFT calculations. Once we establish the foundational data, it can revolutionize the developing process for NNPs and contribute more standard measurements for the model's performance.

|  | Al-FCC (32) $l_b$(Å); $E_{tot}$/atom (eV) | Al-BCC (54) $l_b$(Å); $E_{tot}$/atom (eV) | Al-HCP (48) $l_b$(Å); $E_{tot}$/atom (eV) |
|---|---|---|---|
| Reference (EAM) | 2.86/-3.411 | 2.80/-3.309 | 2.82/-3.380 |
| NNP (Translated Data) | 2.87/-3.407 | 2.82/-3.333 | 2.81/-3.385 |
| (errors) | (+0.01/+0.04) | (+0.02/-0.024) | (-0.01/-0.005) |

Table 3: Results of energy minimization from the different initial structures, FCC, BCC, and HCP aluminum through NNPs from translated data. (blue: lowest energy, red: highest energy, green: middle, $l_b$: bond length)

# Supplementary Information

## Data Distillation for Neural Network Potentials toward Foundational Dataset

Gang Seob Jung[1†], Sangkeun Lee[2] and Jong Youl Choi[2]

[1]Computational Sciences and Engineering Division, Oak Ridge National Laboratory, Oak Ridge, TN 37831, USA
[2]Computational Sciences and Mathematics Division, Oak Ridge National Laboratory, Oak Ridge, TN 37831, USA

†Email: jungg@ornl.gov

**Supplementary Note 1: Data Generation**

*1.1 Molecular dynamics*

The molecular dynamics simulations were performed *via* the LAMMPS package.[1] we utilized EAM potentials for nickel[2] and aluminum.[3] The potentials have been developed to describe liquid/amorphous and crystalline phases. To compare the conventional isobaric-isothermal and MOMT ensemble methods, we prepared a 108 nickel atoms system. We performed a short relaxation through MD simulation at $T_0$ = 2,000 K and $P_0$ = 1 bar for 50,000 steps before sampling. Then, we sampled configurations every 100 steps (0.1 ps with a 1 fs timestep) for 200 ps. The number of obtained data is 2,000 from liquid and FCC phases. Both phases are stable under the conditions based on the initial state, and there is no transition between the two phases through the conventional approach.

*1.2 Multiorder-multithermal ensemble molecular dynamics*

Using the Wang-Landau algorithm,[4] we performed MOMT MD simulation based on the order parameters defined by reciprocal vectors. This method can efficiently sample both liquid and solid phases of crystalline materials, e.g., silicon and MgO.[5,6] Also, thermodynamic quantities as a function of temperature from the reweighting technique were demonstrated in agreement with other estimations in the Lennard-Jones system.[7] In this study, we utilized the method for sampling configurations to evaluate the sampled data for the NNP training. The detailed values for setting nickel and aluminum are listed in Table S1.

In the isobaric-multiorder multithermal (MOMT) ensemble method, the partial enthalpy is defined as functions of $H$ and $O$ as

$$H_{momt} = H_0 + \delta H(H, O)., \tag{1}$$

where $O$ is an order parameter function to evaluate the order of the system as a function of coordinates, $r$, given by[5]

$$O = \frac{1}{N_{fcc} N_A} \sum_{g \in fcc} \left| \sum_i \exp(i g \cdot r_i) \right|^2, \tag{2}$$

where $g$, $N_{fcc}$, and $N_A$ are the shortest reciprocal vectors of FCC, the number of reciprocal vectors, and the number of atoms, respectively. For a more detailed description of methods, we refer to the previous study.[7]

**Supplementary Note 2: Neural Network Potentials**
We utilized the ANI type NNP through TorchANI[8] library. The NN structure in our study is listed in Table S2. We utilized Gaussian error linear unit (GELU) activation function[9]. Atomic environment vectors (AEV), or symmetry function,[10] capture the atomic environmental feature for the NN input. We followed the parameters of the AEV from ANI-2x[11] model except for a longer radius cutoff of 6.9Å with additional Gaussian centers.

For training iteration, 20% was used for validation, and 80% of the data was used to train the model with a mini-batch size 64. The data was shuffled when they were loaded.

The loss function is defined as
$$Loss = \frac{1}{N_{data}} \sum \frac{(E_{NNP}-E_{ref})^2}{\sqrt{N_{atom}}} + \frac{\alpha}{N_{data}} \sum \frac{(\vec{F}_{NNP}-\vec{F}_{ref})^2}{N_{atom}}, \quad (3)$$
where $\alpha$ was set to 0.1, a parameter to determine the contribution of forces. The basic training conditions are the same as in the previous study.[12] The maximum epochs was set as 300 and we took the best parameters for the validation set during the training. We did not prepare an explicit test set. However, we performed structural optimization to check physical properties and evaluated the performance with other data not explicitly included in both train/validation sets.

**Supplementary Note 3: Structural Optimization**
To evaluate the NNPs, we performed the structural optimization for three different closed-pack crystal structures (FCC, BCC, and HCP) and compared the energy rank and bond lengths. Although we sampled the FCC-based solid phase, we did not include the energy-optimized (or at a very low temperature) structure in the training set. The perfectly aligned crystalline structure is less likely to be sampled at a finite temperature, especially near the melting temperature. Whether the trained NNPs can derive the optimized structures can be a valuable indicator of the reliability of NNP models. The optimized shortest bond length and the total energy/atom from the EAM potential are compared using a previously developed interface with LAMMPS[12], we applied the same relaxation process to each structure based on the trained NNPs.

**Supplementary Note 4: Data Distillation**
*3.1 Active learning-based distillation*
Uncertainty quantification (UQ) of the NNPs is a central part of active learning because it allows us to identify valuable data likely to be informative and worth labeling with new calculations. In the current study, we employ the ensemble-based approach, utilizing the same NNP structure but with different training and validation sets for each model.[13] We divided 20,000 data from MOMT sampling into 10 sequential data sets (each set is 2,000 data points). From the initial data set, we trained the NNP with 5 models. At each step, the atomic energy is predicted from each model, and the standard deviation from the model predictions is used as an uncertainty quantification (UQ) measure (atomic UQ). If the atomic UQ value is larger than $\mu_{UQ} + 4\sigma_{UQ}$, we consider the configuration around that atom is not included in the dataset and included in the next iteration.

*3.2 Order parameter & enthalpy based distillation*
For the baseline approach for distillation, we manually selected the data based on the physical properties: the order parameter and enthalpy values. Since the order parameter informs the states of structures, it can be a good indicator of configurational similarity. We utilized data reduction approach through a neighbor list as suggested in the previous study.[12] Table S3 shows the number of data (108 atoms system) selected and deselected based on the δO values.

**Supplementary Note 5: Distilled vs. Non-distilled.**
We checked the training speed and the performance for the structural optimization. As we did in the data distillation, we also trained 5 models for a non-distilled nickel system with 108 atoms. We picked the best model based on their MAE of force. Due to the different numbers of configurations, their training speed is different. We checked the training speed based on 10 epochs in the beginning through one Nvidia GPU in a personnel workstation. For 20,000 data points with a 108-nickel system, it takes 41s/epoch. For 3,500 data points with a 32-nickel system, it takes 2.1s/epoch. The speed-up is about 19.4x. We note that TorchANI provides a CUDA version of symmetry function calculations, but we did not utilize it. Also, the number of batches can affect the benchmark test. The comparison of the two models is shown in Table S6. Full data also results in a good model to predict the energy rank correctly. The sum of the absolute error of bond length indicates that the distilled one is better (Full data: 0.08Å vs Distilled data: 0.03Å). Also, the energy error (max residue of the energy/atom error – min residue of the energy/atom error) shows that the distilled one performed better (Full data: 50meV/atom vs Distilled data: 20meV/atom). It does not completely confirm that distilled one is better, but at least the model trained with the distilled data is comparable with the model with non-distilled data.

Although we utilized empirical potential in the current demonstration, eventually, the applications should be done with DFT calculations. It is inevitable to perform a long-time integration (currently 2 ns) to sample a wide range of configurations in the current study. Since the time-space is not as parallelized as the length-space, running MD for 2ns suddenly becomes impractical with DFT calculations. However, embarrassing parallelization is possible with the sampled configurations (we can ignore time integration). In this context, we expect a clear speed-up, even considering the number of calls to Oracle in the proposed work. Furthermore, the MOMT sampling through NNP is at least two orders of magnitude faster than MOMT sampling through DFT, while the actual speed-up depends on the system size.

**Supplementary Note 6: BCC metal Niobium**
Nickel and Aluminum are FCC-type metals. We further tested whether the distilled data could be translated to BCC-type metal, Niobium (Nb). The parameters for Niobium are available in the empirical potentials we utilized for nickel. We first obtained the relaxed structure of Nb. Since the BCC structure is the most stable, we determined the scaling factor based on BCC as ~1.1867 (2.86/2.41). We did not perform the MOMT MD as the aluminum case, but comparing relaxed structures is still good to show how it would work. We included the results in Table S7. We realized that FCC and HCP of Nb have a very similar energy (2meV/atom difference). Therefore, it is difficult to say that it works perfectly. However, the results are promising, showing similar accuracy with aluminum cases.

**Table S1**. Conditions for the MOMT ensemble molecular dynamics.

| Type (# atoms) | $T$ (K) | $P$ (bars) | $H$ (eV) (# grids) | $O$ (# grids) |
|---|---|---|---|---|
| Ni (108) | 2,000 | 1 | -540 ~ -410 (60) | -5 ~ 108 (100) |
| Ni (32) | 2,000 | 1 | -165 ~ -125 (60) | -5 ~ 40 (110) |
| Al (108) | 1,000 | 1 | -420 ~ -290 (60) | -5 ~ 108 (100) |

**Table S2**. Neural network structures for Ni and Al in the current study. Gaussian error linear unit (GELU) activation function[9] was utilized to add non-linearity between AEV-1$^{st}$, 1$^{st}$-2$^{nd}$, and 2$^{nd}$-3$^{rd}$ layers. The radius cutoff for the radial part is 6.9Å.

| NN Model | 1st | 2nd | 3rd | Output (Energy) |
|---|---|---|---|---|
| Ni/Al/Nb | 224 | 192 | 160 | 1 |

**Table S3**. Data distillation from the 20,000 configurational data sampled by the MOMT ensemble molecular dynamics. Firstly, we trained the 5 NNPs with the selected data (ΔO~3.0). Then, 7 AL iterations were performed from deleted data with a sparse grid (ΔO ~ 3.0) to fine grids (ΔO ~0.02).

| $\Delta O$ | Selected | Deleted | Total |
|---|---|---|---|
| 0.02 | 17,900 | 2,101 | 20,001 |
| 0.05 | 15,516 | 2,384 | 17,900 |
| 0.1 | 12,640 | 2,876 | 15,516 |
| 0.2 | 9,377 | 3,263 | 12,640 |
| 0.4 | 6,234 | 3,143 | 9,377 |
| 0.7 | 4,139 | 2,095 | 6,234 |
| 3.0 | 1,506 | 2,633 | 4,139 |

**Table S4.** The obtained self-energy minimizes the MAE of training/validation sets of each data set sampled from different initial configurations and NPT and MOMT ensembles (108 nickel atoms).

| Data Set | FCC Data | Liquid Data | MOMT Data |
|---|---|---|---|
| Self-Energy (Hartree) | -0.167824160206977 | -0.1610189932429113 | -0.16532494629298 |

**Table S5.** The obtained self-energy minimizes the MAE of training/validation sets of systems

| Data Set | Nickel Data | Aluminum Data | Niobium Data |
|---|---|---|---|
| Self-Energy (Hartree) | -0.16440532763410765 | -0.11764409019342367 | -0.2502205647745999 |

**Table S6.** Results of energy minimization from the different initial structures, FCC, BCC, and HCP nickel through NNPs from distilled data and non-distilled data (Full data: 20,000 data points of 108 atoms). (blue: lowest energy, red: highest energy, green: middle, $l_b$: bond length).

| Structure (#atoms) | Ni-FCC (32) $l_b$(Å); $E_{tot}$/atom (eV) | Ni-BCC (54) $l_b$(Å); $E_{tot}$/atom (eV) | Ni-HCP (48) $l_b$(Å); $E_{tot}$/atom (eV) |
|---|---|---|---|
| Reference (EAM) | 2.49/-4.876 | 2.41/-4.833 | 2.44/-4.847 |
| NNP (Full Data) (errors) | 2.46/-4.896 (-0.03/-0.020) | 2.45/-4.803 (+0.04/+0.030) | 2.43/-4.848 (-0.01/-0.001) |
| NNP (Distilled Data) (errors) | 2.50/-4.875 (+0.01/+0.001) | 2.41/-4.812 (0.00/+0.021) | 2.46/-4.835 (+0.02/0.012) |

**Table S7.** Results of energy minimization from the different initial structures, FCC, BCC, and HCP Niobium (Nb) through NNPs from translated data. (blue: lowest energy, red: highest energy, green: middle, $l_b$: bond length)

| Structure (#atoms) | Nb-FCC (32) $l_b$(Å); $E_{tot}$/atom (eV) | Nb-BCC (54) $l_b$(Å); $E_{tot}$/atom (eV) | Nb-HCP (48) $l_b$(Å); $E_{tot}$/atom (eV) |
|---|---|---|---|
| Reference (EAM) | 3.05/-7.159 | 2.86/-7.347 | 2.94/-7.157 |
| NNP (Translated Data) (errors) | 3.07/-7.173 (0.02/-0.014) | 2.86/-7.350 (+0.00/-0.003) | 2.93/-7.166 (-0.01/-0.009) |

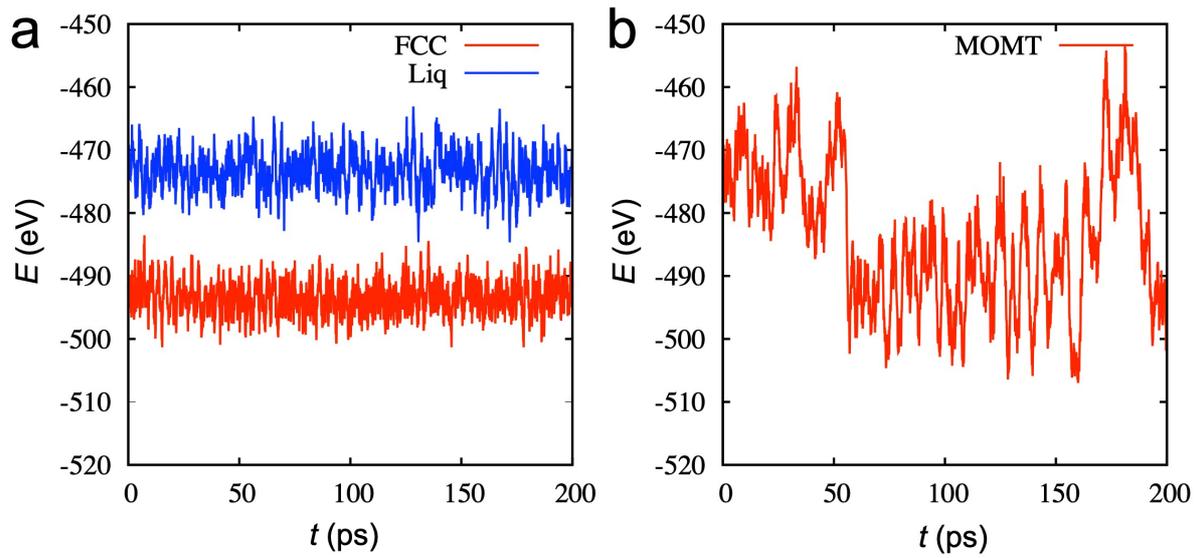

**Figure S1.** Time evolutions of energy ($E$) of nickel at $T_0$=2,000 K, $P_0$=1.0 bar for 200 ps. (**a**) Conventional NPT ensemble with different initial phases, FCC, and liquid phases. (**b**) Time evolution of E through MOMT ensemble. It shows that the transition between the two phases and the energy range where the liquid and FCC phases can be efficiently sampled.

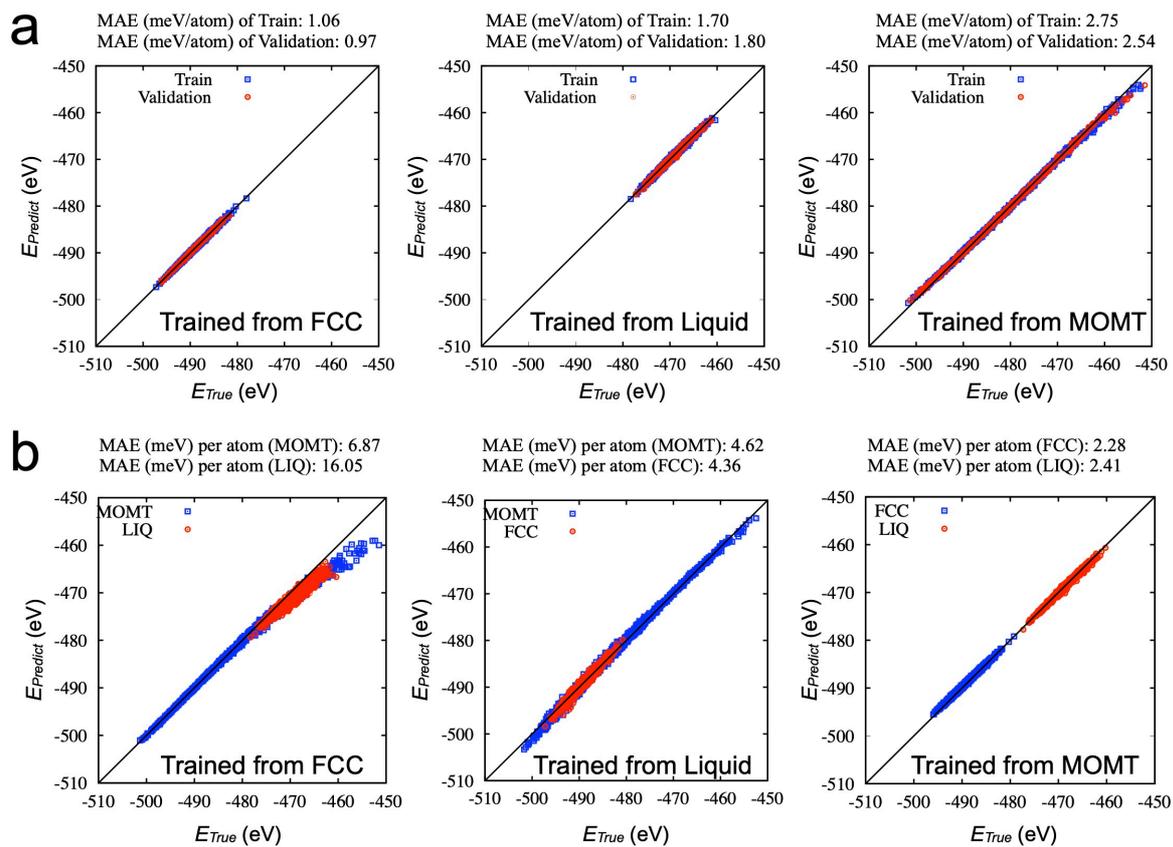

**Figure S2** Mean absolute error (MAE) of NNPs trained from three different sampled data. (**a**) MAE of training/validation sets (**b**) MAE of other data (unseen) sets for each NNP model.

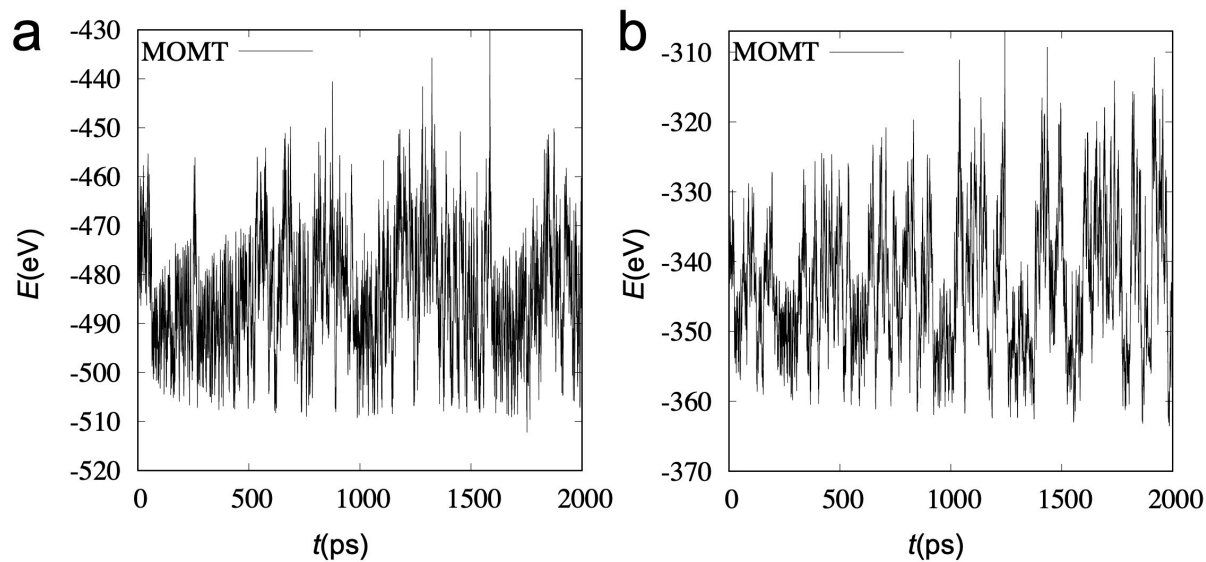

**Figure S3** Times series of the MOMT ensemble MD simulations: (**a**) 108 nickel atoms at 2,000 K and 1 bar (**b**) 108 aluminum atoms at 1,000 K and 1 bar. The sampling region can change due to the reference temperature and maximum partial enthalpy. The range of the $y$ label is set by scaling the $y$ range of nickel. In the aluminum case, it can sample higher energy regions than nickel.